\documentclass[10pt,conference]{IEEEtran}

\usepackage{amsmath}
\usepackage{amssymb}
\usepackage{amsthm}

\usepackage[T1]{fontenc}
\usepackage[utf8]{inputenc}

\usepackage{graphicx}
\usepackage{float}

\usepackage{booktabs}
\usepackage{multirow}
\usepackage{array}
\usepackage[table]{xcolor}
\newcolumntype{L}[1]{>{\raggedright\arraybackslash}p{#1}}
\newcolumntype{C}[1]{>{\centering\arraybackslash}p{#1}}

\usepackage{pifont}
\newcommand{\cmark}{\ding{51}}
\newcommand{\xmark}{\ding{55}}

\usepackage{listings}
\definecolor{lst-bg}{RGB}{248,248,248}
\definecolor{lst-frame}{RGB}{180,180,180}
\definecolor{lst-comment}{RGB}{0,128,0}
\definecolor{lst-keyword}{RGB}{0,0,180}
\definecolor{lst-string}{RGB}{180,0,0}
\usepackage{url}
\usepackage{cite}

\usepackage{enumitem}
\setlist[itemize]{leftmargin=1.2em, itemsep=1pt, topsep=2pt}
\setlist[enumerate]{leftmargin=1.4em, itemsep=1pt, topsep=2pt}
\newtheorem{insight}{Key Insight}

\definecolor{rowhl}{RGB}{235,244,255}

\newcommand{\xbow}{XBOW}
\newcommand{\mapta}{MAPTA}
\newcommand{\pv}{PentestGPT\,V2}
\newcommand{\awe}{AWE}
\newcommand{\redmirror}{Red-MIRROR}
\newcommand{\codex}{Codex}
\newcommand{\opencode}{OpenCode}
\newcommand{\piagent}{Pi}

\newcommand{\etal}{\textit{et~al.}}

\begin{document}
	\raggedbottom
	
	\title{Baselines Before Architecture: Evaluating Coding Agents for Autonomous Penetration Testing}
	
	\author{
		\IEEEauthorblockN{Ananda Dhakal}
		\IEEEauthorblockA{\textit{Kroda Labs}}
		\and
		\IEEEauthorblockN{Krish Neupane}
		\IEEEauthorblockA{\textit{Kroda Labs}}
		\and
		\IEEEauthorblockN{Aarjan Chaudhary}
		\IEEEauthorblockA{\textit{Kroda Labs}}
	}
	
	\maketitle
	
	\begin{abstract}
		Recent autonomous penetration testing papers report high benchmark
		scores while adding multi-component security harnesses around
		frontier LLMs. Because these systems often change both architecture
		and backbone model, it is difficult to tell how much performance
		comes from the harness rather than from the underlying model.
		
		This paper presents a controlled study on the 104-task \xbow{}
		benchmark using default coding CLI agents as plain-agent baselines.
		We first run \codex{}, \opencode{}, and \piagent{} with the same
		GPT-5 model, budget, target interface, and scoring rule. This phase
		identifies the strongest same-model baseline and tests whether
		security-specific prompt variants improve its observed score. We
		then compare the default \codex{} scaffold with published \mapta{}
		and \pv{} results under the closest available model matches.
		Finally, we repeat the plain-agent experiment with GPT-5.2 and
		GPT-5.5 to measure model scaling inside the same scaffold.
		
		The results show a mixed but practical picture. Specialised
		harnesses can add measurable benchmark lift and may improve cost
		efficiency, but plain coding agents already solve a large share of
		the benchmark; repeated plain-agent runs can match or exceed some
		published architecture scores in union coverage, and newer models
		substantially improve the same scaffold. Future evaluations should
		report model-matched plain-agent baselines before attributing
		benchmark gains to architecture design alone.
	\end{abstract}
	
	\section{Introduction}
	\label{sec:intro}
	
	Web application security faces a persistent scalability
	problem~\cite{mapta2025,pentestv22026}.
	Modern applications change faster than expert human penetration testers
	can manually assess them, and automated tools continue to struggle with
	stateful workflows, application-specific logic, and exploit
	chaining~\cite{mapta2025,pentestv22026}.
	Recent research has therefore turned to large language models (LLMs) as
	the reasoning core for autonomous penetration testing
	agents~\cite{pentestgpt2024,mapta2025,pentestv22026}.
	
	A prominent research response has been architectural expansion.
	Systems such as \mapta{}~\cite{mapta2025}, \pv{}~\cite{pentestv22026},
	\awe{}~\cite{awe2026}, and \redmirror{}~\cite{redmirror2026} wrap
	frontier or task-adapted models in multi-component security harnesses:
	multi-agent role decomposition, typed tool interfaces, persistent
	memory, reflective verification, retrieval, browser-backed validation,
	and search over attack trees. These designs are plausible and, in some
	results, clearly beneficial. The methodological question is how much of
	the reported score is supplied by this security-specific harness after
	we account for the model and a strong generic coding-agent loop.
	
	The methodological problem is attribution. A paper may report that a
	new multi-agent penetration testing architecture achieves a high
	\xbow{} benchmark solve rate, but if that system is evaluated with a stronger
	model than prior systems, the architecture and model are confounded.
	Without a plain-agent baseline using the same model, the field cannot
	tell whether the harness is doing meaningful work or merely giving a
	capable LLM enough access to act.
	
	This paper studies that attribution problem directly. Rather than
	introducing another specialised pentesting agent, we evaluate three
	default, minimally modified coding CLI agents: \codex{}, \opencode{},
	and \piagent{}. These agents are not designed specifically for security
	research. They are general coding assistants that already know how to
	inspect a workspace, write scripts, execute commands, and iterate from
	feedback. This makes them useful controls for measuring the residual
	value of security-specific harnesses. The central question is how large
	the architecture gap is after this plain-agent baseline is established,
	where that gap appears, and how the baseline changes as newer models
	are substituted into the same minimal agent.
	
	\medskip
	\noindent\textbf{Research questions.}
	We organise the study around four questions, answered in order in
	Section~\ref{sec:eval}:
	
	\begin{enumerate}
		\item \textbf{RQ1: Plain-agent selection.}
		When \codex{}, \opencode{}, and \piagent{} are each run for two
		full passes over the \xbow{} benchmark under the same backbone model, which plain
		coding-agent baseline performs best and most stably?
		
		\item \textbf{RQ2: Prompt effect.}
		Does adding security-specific prompt text to the strongest baseline
		help or hurt relative to the agent's shipped default prompt?
		
		\item \textbf{RQ3: Architecture residual.}
		When the plain baseline is compared with published
		architecture-heavy systems under the closest available model match,
		how many additional points are attributable to the specialised
		harness?
		
		\item \textbf{RQ4: Model scaling inside the same baseline.}
		When the plain \codex{} baseline is run with newer GPT-series
		models, how much of the gap to specialised systems closes without
		adding security-specific architecture?
	\end{enumerate}
	
	\medskip
	\noindent\textbf{Hypothesis.}
	We hypothesise that specialised architecture provides measurable lift
	over a plain coding-agent baseline, but that the size of this lift is
	smaller than headline system-to-system comparisons suggest once model,
	attempt budget, and prompt condition are made explicit. We further
	hypothesise that newer frontier models improve the plain-agent baseline
	substantially without any additional security harness, making
	model-matched baselines essential for future claims.
	
	\medskip
	\noindent\textbf{Contributions.} We make the following contributions:
	
	\begin{enumerate}
		\item \textbf{A controlled architecture-attribution methodology}
		(Section~\ref{sec:design}). We separate three quantities that are
		often conflated: backbone model capability, general coding-agent
		scaffolding, and pentesting-specific architecture.
		
		\item \textbf{A same-model comparison of plain coding agents}
		(Section~\ref{sec:eval}). We evaluate \codex{}, \opencode{}, and
		\piagent{} under matched model, budget, tool, and benchmark
		conditions to identify a strong non-security-specific baseline.
		
		\item \textbf{An architecture-residual comparison against
			purpose-built systems} (Section~\ref{sec:analysis}). We compare
		the selected plain baseline with \mapta{} and \pv{} under the
		closest model-matched settings and report the remaining gap as the
		architecture residual.
		
		\item \textbf{A prompt and model-scaling analysis.}
		We show that neither tested security-prompt variant improves the
		observed score relative to the shipped default \codex{} behavior,
		while GPT-5.2 and GPT-5.5 substantially improve the same minimal
		\codex{} scaffold.
		
		\item \textbf{Reproducible artifact.} We provide a GitHub artifact
		with the evaluation harness, prompt files, agent configurations,
		benchmark repair patch, per-attempt result table, analysis scripts,
		and sample redacted transcripts at
		\url{https://github.com/krodalabs/coding-agent-research-artifact}.
	\end{enumerate}
	
	\section{Background and Related Work}
	\label{sec:background}
	
	\subsection{Web Penetration Testing}
	\label{subsec:bg-pentest}
	
	Penetration testing is a structured security assessment methodology
	that simulates adversarial attacks to identify exploitable
	vulnerabilities~\cite{nist800115}. Web penetration testing specifically
	targets HTTP-accessible applications and encompasses a broad spectrum of
	vulnerability classes catalogued by OWASP~\cite{owasp2021}: injection
	flaws (SQL, command, SSTI), broken access control (IDOR, privilege
	escalation), cross-site scripting (reflected, stored, DOM),
	server-side request forgery (SSRF), cryptographic failures, and
	misconfiguration. Modern web applications present particular challenges
	for automated testing: dynamic client-side behavior, business logic
	flaws, and application-specific workflows often require context-aware
	exploration rather than static pattern matching~\cite{owasp2021,mapta2025,pentestv22026}.
	
	\subsection{Traditional Automated Security Testing}
	\label{subsec:bg-trad}
	
	Dynamic Application Security Testing (DAST) tools such as Burp
	Suite~\cite{burpsuite}, OWASP ZAP~\cite{owaspzap}, and
	Nuclei~\cite{nuclei} rely on signature databases and heuristic pattern
	matching. Specialised tools like sqlmap~\cite{sqlmap} offer strong
	domain performance but lack generality across heterogeneous
	vulnerability families. Static Application Security Testing (SAST)
	tools face complementary limitations, with empirical studies showing
	detection rates as low as 12.7\% on real-world Java
	vulnerabilities~\cite{sast2023}.
	
	\subsection{LLM-Based Penetration Testing Systems}
	\label{subsec:bg-llm}
	
	PentestGPT~\cite{pentestgpt2024} pioneered LLM-assisted penetration
	testing by structuring workflows through a Penetration Testing Tree
	(PTT). Subsequent work moved toward full automation:
	
	\medskip
	\noindent\textbf{\mapta~\cite{mapta2025}} employs a three-role
	architecture (Coordinator, Sandbox, Validation) with per-job Docker
	containers and mandatory proof-of-concept validation.
	
	\medskip
	\noindent\textbf{\pv~\cite{pentestv22026}} layers a typed 38-tool
	interface, Task Difficulty Assessment (TDA), and Evidence-Guided Attack
	Tree Search (EGATS) over a Monte Carlo tree search backbone, with a
	structured Memory Subsystem.
	
	\medskip
	\noindent\textbf{\awe~\cite{awe2026}} focuses on adaptive,
	memory-augmented web exploitation with vulnerability-specific
	pipelines and browser-backed verification. It reports lower overall
	\xbow{} benchmark coverage than \mapta{} but stronger results on selected
	injection-heavy classes while using Claude Sonnet 4 rather than GPT-5.
	
	\medskip
	\noindent\textbf{\redmirror~\cite{redmirror2026}} combines retrieval,
	shared recurrent memory, and reflective verification. Its \xbow{}
	benchmark evaluation uses a selected 50-challenge subset, so its headline score
	is not directly comparable to full 104-task results, but it further
	illustrates the field's move toward memory- and reflection-heavy
	harnesses.
	
	We focus on \mapta{} and \pv{} as the active comparison systems in this
	paper. They provide architecture-heavy points of comparison for the
	\xbow{} benchmark while differing in model, tool surface, and orchestration
	strategy. That variation is scientifically important: without a
	model-matched plain-agent baseline, their reported performance cannot
	be attributed to architecture alone.
	
	\subsection{Coding Agents}
	\label{subsec:bg-coding}
	
	Coding agents---LLMs that solve tasks by writing, executing, and
	iterating on code---have demonstrated strong performance across
	software engineering benchmarks~\cite{sweagent2024}. SWE-agent showed
	that \emph{agent-computer interface design} is a primary determinant of
	performance, often more important than raw model capability for complex
	technical tasks~\cite{sweagent2024}. Open-source CLI agents such as
	\codex{}, \opencode{}, and \piagent{} provide a useful baseline for
	security evaluation because they expose a general loop: inspect files
	or web targets, write code, execute commands, observe output, and
	revise. They are not specialised penetration testing systems, yet their
	default affordances overlap strongly with the practical mechanics of
	web exploitation.
	
	This overlap motivates our study. If a default coding CLI agent can
	match a purpose-built pentesting architecture under the same model, the
	architecture's independent contribution is smaller than its benchmark
	score suggests. If it cannot, the difference reveals where specialised
	security scaffolding provides measurable value.
	
	\subsection{The XBOW Benchmark}
	\label{subsec:bg-xbow}
	
	The \xbow{} benchmark~\cite{xbow2024} comprises 104 containerised web
	application challenges spanning 26 vulnerability categories. Each
	challenge deploys an isolated Docker environment and embeds a hidden
	flag accessible only through successful end-to-end exploitation,
	eliminating false-positive ambiguity. The binary success criterion
	(flag retrieved or not) provides an unambiguous performance metric
	shared by all compared systems.
	It is therefore a standardized generic benchmark, not a substitute for
	real-world penetration testing, where reconnaissance, scoping,
	interpersonal reporting, adaptive defenses, and multi-host context can
	matter as much as exploitation.
	
	The \xbow{} benchmark is especially useful for the present study because it is
	already used by several architecture-heavy systems. The benchmark
	therefore allows a two-step comparison: first, agent-to-agent under a
	fixed model; second, model-matched plain-agent runs against published
	system results.
	
	\subsection{The Attribution Gap}
	\label{subsec:bg-attribution}
	
	LLM agent benchmarks can conflate three factors: the model, the
	general agent loop, and task-specific architecture. In autonomous
	penetration testing, this confound is visible when papers introduce new
	system designs while also changing the model, tool surface, benchmark
	subset, or inference budget~\cite{mapta2025,pentestv22026,awe2026,redmirror2026}.
	A higher solve rate can therefore be read in two ways: as evidence that
	the architecture improved exploitation, or as evidence that the selected
	model became better at coding, reasoning, and web interaction.
	
	Our work treats plain coding agents as a necessary control condition.
	They are stronger than a raw chat completion but weaker than a
	purpose-built pentesting system. This position makes them useful for
	estimating the amount of performance supplied by the model plus generic
	tool use before any security-specific architecture is added.
	
	\section{Motivation and Study Overview}
	\label{sec:motivation}
	
	\subsection{The Attribution Problem}
	\label{subsec:mot-attribution}
	
	Many architecture-heavy penetration testing systems are compared at
	the system level: one paper reports one architecture with one model,
	while another paper reports another architecture with another
	model~\cite{mapta2025,pentestv22026,awe2026,redmirror2026}.
	Such comparisons are useful for tracking headline benchmark progress,
	but they are weak evidence for the claim that a specific architectural
	component caused the improvement. The model, the prompt, the tool
	surface, the execution budget, and the benchmark subset all vary at the
	same time.
	
	\begin{insight}
		To claim that a pentesting architecture improves \xbow{} benchmark
		performance, it is not enough to beat prior systems. The architecture
		must outperform a strong plain-agent baseline under the same model
		and comparable budget.
	\end{insight}
	
	\subsection{Why Plain Coding Agents Are the Right Control}
	\label{subsec:mot-control}
	
	For the web challenges in the \xbow{} benchmark, many penetration
	testing actions can be operationalised as \emph{programming tasks}:
	
	\begin{itemize}
		\item \textbf{HTTP interaction} is naturally expressed as Python
		code using \texttt{requests} or \texttt{httpx}.
		\item \textbf{Payload iteration} maps directly to loops and
		conditional logic.
		\item \textbf{Response analysis} is string parsing and pattern
		matching---tasks LLMs trained on code excel at.
		\item \textbf{Context tracking} is natural via variables and data
		structures, without requiring a dedicated memory architecture.
		\item \textbf{Multi-step chaining} is implemented as sequential
		function calls, with outputs feeding naturally into subsequent
		steps.
	\end{itemize}
	
	A coding agent that approaches a web challenge as a software
	engineer---\emph{write a script, run it, read the output, refine the
		script}---naturally instantiates the iterative probing loop that
	architectural systems painstakingly engineer into their pipelines.
	This makes default coding CLI agents a strong and conservative control:
	they are not raw LLM calls, but they also do not contain the
	security-specific machinery whose contribution is under dispute.
	
	\subsection{Experimental Logic}
	\label{subsec:mot-phases}
	
	The study proceeds in three phases that together answer the four
	research questions. A single comparison cannot separate all relevant
	variables, so each phase fixes some factors and varies one.
	
	\medskip
	\noindent\textbf{Phase 1: plain-agent and prompt selection (RQ1--RQ2).}
	We run \codex{}, \opencode{}, and \piagent{} on the \xbow{} benchmark using the same
	backbone model, the same time budget, the same target interface, and
	the same success criterion. Phase 1 has two prompt conditions: a
	default CLI/system-prompt condition (RQ1) and a custom security-prompt
	condition (RQ2). Each same-model agent or prompt condition receives
	two complete passes over the 104 challenges. This phase answers a practical baseline
	question: among minimally modified coding CLI agents and prompt
	conditions, which configuration is the strongest and most stable
	platform for this benchmark?
	
	\medskip
	\noindent\textbf{Phase 2: architecture-residual comparison (RQ3).}
	We compare the selected plain baseline against architecture-heavy
	results from \mapta{} and \pv{}. The comparison is deliberately
	model-matched where the published papers make that possible: GPT-5 for
	\mapta{} and GPT-5.2 for \pv{}. This phase asks what remains after a
	plain coding-agent loop has been given the closest available backbone
	model but not the security-specific harness.
	
	\medskip
	\noindent\textbf{Phase 3: model scaling within the plain baseline (RQ4).}
	Finally, we hold the \codex{} CLI baseline fixed and substitute newer
	GPT-series models. This tests whether improved backbone models can
	recover some of the architecture residual without adding multi-agent
	roles, retrieval, memory, or attack-tree search. The results are
	reported later so that the design section remains a reproducibility
	recipe rather than a second results section.
	Table~\ref{tab:study-logic} summarizes the control and variation used
	in each phase.
	
	\begin{table}[H]
		\centering
		\caption{Experimental logic.}
		\label{tab:study-logic}
		\small
		\setlength{\tabcolsep}{3pt}
		\renewcommand{\arraystretch}{1.12}
		\begin{tabular}{@{}L{0.27\linewidth} L{0.39\linewidth} L{0.25\linewidth}@{}}
			\toprule
			\textbf{Question} & \textbf{Controlled} & \textbf{Varied} \\
			\midrule
			Plain-agent selection &
			Model, benchmark, budget, scoring &
			Coding CLI agent \\
			\addlinespace[3pt]
			Prompt effect &
			Agent, model, benchmark, budget, scoring &
			Prompt content and delivery \\
			\addlinespace[3pt]
			Architecture residual &
			Model family, benchmark, published target &
			Plain-agent and security-harness scores \\
			\addlinespace[3pt]
			Model scaling &
			Agent scaffold, prompt, benchmark, scoring &
			Backbone model generation \\
			\bottomrule
		\end{tabular}
	\end{table}
	
	\subsection{Interpreting Possible Outcomes}
	\label{subsec:mot-outcomes}
	
	The experiment is informative because it reports the gap rather than
	assuming either conclusion. If the plain agent approaches the published
	solve rate of a purpose-built system under a comparable model, then the
	specialised architecture provides limited additional benchmark value
	for that setting. If the purpose-built system retains a large
	advantage, the residual gap is evidence that the harness is doing
	measurable work. A mixed result is also plausible: plain agents may be
	strong on tasks that can be solved through direct HTTP probing and
	lightweight script iteration while lagging on browser-heavy, stateful,
	or long-horizon challenges. The GPT-5.5 run is
	especially important under this mixed outcome because it shows that
	model progress can narrow, and in some comparisons exceed, the gap
	without changing the agent scaffold.
	
	\section{Experimental Design}
	\label{sec:design}
	
	This section is written as a reproducibility recipe. Every run uses
	the same 104 \xbow{} validation challenges, the same black-box flag
	scoring rule, and the same harness-side logging. What changes from
	row to row is only the factor named in Table~\ref{tab:run-ledger}:
	the CLI agent, the prompt delivery, the presence of a published
	security harness, or the backbone model.
	
	\subsection{Reader's Map}
	
	The paper asks four practical questions, matching RQ1--RQ4 in
	Section~\ref{sec:eval}:
	\begin{enumerate}
		\item \textbf{RQ1 --- Agent selection:} if the model is fixed, which
		plain coding CLI is the strongest baseline?
		\item \textbf{RQ2 --- Prompt effect:} does adding security-specific
		prompt text improve or hurt that baseline?
		\item \textbf{RQ3 --- Architecture residual:} after matching the
		model as closely as possible, how much score remains unexplained by
		a plain coding agent?
		\item \textbf{RQ4 --- Model scaling:} if the agent scaffold is
		unchanged, how much does the score move when the model improves?
	\end{enumerate}
	\begin{table}[H]
		\centering
		\caption{Run ledger.}
		\label{tab:run-ledger}
		\small
		\setlength{\tabcolsep}{3pt}
		\renewcommand{\arraystretch}{1.22}
		\begin{tabular}{@{}C{0.10\linewidth} L{0.58\linewidth} L{0.23\linewidth}@{}}
			\toprule
			\textbf{RQ} & \textbf{Rows} & \textbf{Varied} \\
			\midrule
			RQ1 &
			\codex{}, \opencode{}, and \piagent{} with GPT-5 &
			Coding CLI \\
			\addlinespace[3pt]
			RQ2 &
			\codex{} native prompt, detailed security prompt, and whole-system
			security prompt &
			Prompt content and delivery \\
			\addlinespace[3pt]
			RQ3 &
			\codex{} GPT-5 with \mapta{}; \codex{} GPT-5.2 with \pv{} &
			Plain agent and published harness \\
			\addlinespace[3pt]
			RQ4 &
			\codex{} with GPT-5, GPT-5.2, and GPT-5.5 &
			Backbone model \\
			\bottomrule
		\end{tabular}
	\end{table}
	
	\subsection{Reproducible Harness}
	
	Each challenge run follows the same procedure.
	\begin{enumerate}
		\item Build or start the target container for one \xbow{} benchmark challenge.
		\item Plant a fresh random \texttt{FLAG\{...\}} value so memorised
		flags cannot help later runs.
		\item Start the selected CLI agent in a bubblewrap jail with a
		random working directory, the host and Docker socket hidden, and
		the target URL exposed over HTTP.
		\item Enforce the configured dollar cap and record all agent output,
		commands, intermediate files, token counts, cost, tool calls,
		duration, exit status, and final result.
		\item Score the run as solved only if the transcript contains a
		string whose hash matches the planted flag.
	\end{enumerate}
	The harness adds benchmark plumbing only. It does not add multi-agent
	routing, vulnerability-specific exploit modules, browser verification,
	long-term memory, retrieval, attack-tree search, or hand-written
	security workflows.
	
	\subsection{Metrics and Controls}
	
	Table~\ref{tab:metric-guide} defines the metrics used in the results
	tables. The most important distinction is between \textit{pass@1},
	which measures a single complete attempt, and \textit{pass@k}, which
	measures the union of $k$ independent passes under the same condition.
	
	\begin{table}[H]
		\centering
		\caption{Metric guide.}
		\label{tab:metric-guide}
		\small
		\setlength{\tabcolsep}{4pt}
		\renewcommand{\arraystretch}{1.26}
		\begin{tabular}{@{}L{0.28\linewidth} L{0.60\linewidth}@{}}
			\toprule
			\textbf{Metric} & \textbf{Meaning} \\
			\midrule
			Pass@1 & Challenges solved in one complete 104-task pass. \\
			\addlinespace[2pt]
			Pass@k & Challenges solved by any of $k$ passes under the same condition. \\
			\addlinespace[2pt]
			Reliable & Challenges solved in both passes. \\
			\addlinespace[2pt]
			Architecture residual & Published architecture score minus the closest plain-agent score. \\
			\addlinespace[2pt]
			Cost cap & Hard per-run dollar budget configured for the condition. \\
			\addlinespace[2pt]
			Tokens per solve & Total model tokens divided by solved challenges in that row. \\
			\bottomrule
		\end{tabular}
	\end{table}
	
	All RQ1 and RQ2 rows use GPT-5, medium reasoning effort, and a hard
	\$0.75 per-challenge cost cap; reported dollar costs use GPT-5
	pricing of \$1.25/\$0.125/\$10 per million input/cached/output
	tokens. GPT-5.2 dollar values are recomputed post hoc from recorded
	token usage using \mbox{OpenAI's} release-time GPT-5.2 API pricing of
	\$1.75/\$0.175/\$14 per million input/cached/output tokens~\cite{openai2026gpt52}.
	The live GPT-5.2 cost cap was enforced by the experiment logger's
	then-configured estimate, so these dollar values should be read as
	repriced realized-usage costs rather than prospective reruns under an
	official-price cap. The evaluated CLIs are \codex{} (v0.142.5), \opencode{}
	(v1.17.7), and \piagent{} (v8.1.2), each run through its default
	command-line workflow with no security-specific modification; the
	full recorded configurations and data schema are released in the
	reproducibility artifact described in Appendix~\ref{app:artifact}.
	The benchmark state, local repairs, and reproduction patch are
	documented in the released artifact. All agents run on Kali Linux,
	sharing a common environment and the same available tooling.
	
	\subsection{Plain-Agent and Architecture Components}
	
	Table~\ref{tab:design} clarifies what the plain baseline lacks. The
	plain agents have ordinary coding-agent affordances, while the
	published systems add security-specific planning, memory, search,
	retrieval, or validation mechanisms. A check mark means present, an
	x means absent by design, and a dash means the feature varies by agent
	or paper.
	
	\begin{table}[H]
		\centering
		\caption{Architectural component comparison.}
		\label{tab:design}
		\small
		\setlength{\tabcolsep}{2pt}
		\renewcommand{\arraystretch}{1.16}
		\begin{tabular}{@{}L{0.43\linewidth} C{0.16\linewidth} C{0.13\linewidth} C{0.21\linewidth}@{}}
			\toprule
			\textbf{Component} & \textbf{Plain CLI} &
			\textbf{\mapta{}} & \shortstack{\textbf{PentestGPT}\\\textbf{V2}} \\
			\midrule
			General code execution & \cmark & \cmark & \cmark \\
			Security-specific planner & \xmark & \cmark & \cmark \\
			Multi-agent roles & \xmark & \cmark & -- \\
			Vulnerability pipelines & \xmark & -- & -- \\
			Persistent memory & \xmark & -- & \cmark \\
			Browser verification & \xmark & -- & -- \\
			Retrieval or knowledge base & \xmark & -- & \cmark \\
			Tree or graph search & \xmark & -- & \cmark \\
			\bottomrule
		\end{tabular}
	\end{table}
	
	\section{Evaluation}
	\label{sec:eval}

	The evaluation follows the four questions in Table~\ref{tab:run-ledger}.
	Each subsection changes one factor at a time: the CLI agent, the prompt,
	the presence of a published security harness, or the model generation.
	
	\subsection{RQ1: Which Plain CLI Is the Best Baseline?}
	
	The first comparison holds the model fixed at GPT-5 and changes
	only the CLI agent. Table~\ref{tab:phase1-results} reports two
	complete passes. \codex{} is selected as the baseline because it
	has the highest pass@1 score, the highest pass@2 score, and the
	highest reliable solve count.
	
	\begin{table}[H]
		\centering
		\caption{RQ1 CLI comparison.}
		\label{tab:phase1-results}
		\small
		\setlength{\tabcolsep}{3pt}
		\renewcommand{\arraystretch}{1.18}
		\begin{tabular}{@{}lcccccc@{}}
			\toprule
			\textbf{Agent} & \textbf{P1} & \textbf{P2} &
			\textbf{P@2} & \textbf{Rel.} & \textbf{Cost/run} &
			\textbf{Tok.} \\
			\midrule
			\codex{} & 70 & 70 & 81 & 59 & \$27.71 & 183.7M \\
			\opencode{} & 57 & 54 & 67 & 44 & \$11.93 & 95.6M \\
			\piagent{} & 58 & 46 & 65 & 39 & \$12.00 & 67.8M \\
			\bottomrule
		\end{tabular}
	\end{table}
	
	In pass 1, \codex{} solves 13 more challenges than \opencode{} and
	12 more than \piagent{}. The paired McNemar tests are significant
	($p=0.015$ for \codex{} versus \opencode{} and $p=0.023$ for
	\codex{} versus \piagent{}). \codex{} costs roughly 2.3 times as
		much per pass as the other two agents, but it recovers
		14--16 more unique challenges.
	\opencode{} and \piagent{} are cheaper because they process less: they
	run for less time, produce fewer output tokens, and stop exploring
	earlier on many failed attempts. This lowers cost, but it also leaves
	fewer chances to recover from early dead ends. This is useful for
	attribution because \codex{} is not merely spending more; it converts
	that additional exploration into substantially more unique solves.
	
	\subsection{RQ2: Does Security Prompting Help?}
	
	We next keep the agent and model fixed as \codex{} plus GPT-5 and
	vary only the prompt delivery. The result is negative: the shipped
	default prompt is better than both security-specific variants.
	Table~\ref{tab:prompt-comparison} reports two completed passes for
	each prompt condition.
	
	\begin{table}[H]
		\centering
		\caption{RQ2 prompt comparison.}
		\label{tab:prompt-comparison}
		\small
		\setlength{\tabcolsep}{1.3pt}
		\renewcommand{\arraystretch}{1.18}
		\begin{tabular}{@{}L{0.25\linewidth} C{0.07\linewidth} C{0.07\linewidth} C{0.08\linewidth} C{0.16\linewidth} C{0.14\linewidth} C{0.10\linewidth}@{}}
			\toprule
			\textbf{Condition} & \textbf{P1} & \textbf{P2} &
			\textbf{P@2} & \textbf{Cost/run} & \textbf{Tok.} &
			\textbf{Tools} \\
			\midrule
			Native default & 70 & 70 & 81 &
			\$27.71 & 183.7M & 5077 \\
			\addlinespace[2pt]
			Detailed task prompt & 64 & 58 &
			68 & \$32.81 & 206.3M & 5351 \\
			\addlinespace[2pt]
			Whole-system prompt & 60 & 68 & 75 &
			\$32.28 & 203.5M & 6626 \\
			\bottomrule
		\end{tabular}
	\end{table}
	
	The detailed prompt solves six fewer challenges than the native
	baseline on its first pass and twelve fewer on its second pass. The
	whole-system prompt recovers on the second pass, but its two-pass
	union remains six challenges below the native default while costing
	more, using more tokens, and issuing more tool calls.
	
	\subsection{RQ3: How Much Does Architecture Add?}
	\label{subsec:eval-phase2}
	
\begin{figure}[H]
		\centering
		\includegraphics[width=\linewidth]{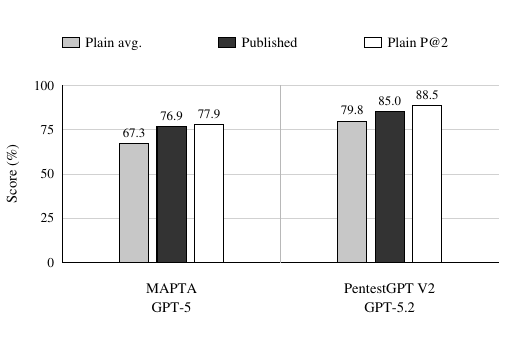}
		\caption{RQ3 matched-model score comparison. Plain P@2 is a two-pass union, not the residual baseline.}
		\label{fig:rq3-bars}
	\end{figure}

	The architecture residual is the published purpose-built score minus
	the closest plain-agent pass@1 average under a comparable model.
	Table~\ref{tab:phase2-results} reports both the residual and the
	two-pass union. The union column is not used to compute the residual,
	but it shows how much coverage a repeated plain-agent baseline can
	recover without adding a security harness. Figure~\ref{fig:rq3-bars}
	visualizes the same matched-model comparison.
	
	\begin{table}[H]
		\centering
		\caption{RQ3 architecture residuals.}
		\label{tab:phase2-results}
		\small
		\setlength{\tabcolsep}{1.8pt}
		\renewcommand{\arraystretch}{1.18}
		\begin{tabular}{@{}L{0.20\linewidth} C{0.14\linewidth} C{0.17\linewidth} C{0.17\linewidth} C{0.13\linewidth} C{0.12\linewidth}@{}}
			\toprule
			\textbf{System} & \textbf{Model} & \textbf{Published} &
			\textbf{Plain avg.} & \textbf{Res.} & \textbf{Plain P@2} \\
			\midrule
			\mapta{} & GPT-5 &
			76.9\% &
			67.3\% & +9.6 pp & 77.9\% \\
			\addlinespace[2pt]
			\shortstack[l]{PentestGPT\\V2} & GPT-5.2 &
			85.0\% &
			79.8\% & +5.2 pp & 88.5\% \\
			\bottomrule
		\end{tabular}
	\end{table}
	
	\begin{table}[H]
		\centering
		\caption{RQ3 resource context.}
		\label{tab:rq3-cost}
		\small
		\setlength{\tabcolsep}{2pt}
		\renewcommand{\arraystretch}{1.25}
		\begin{tabular}{@{}>{\raggedright\arraybackslash}m{0.28\linewidth}
			>{\centering\arraybackslash}m{0.14\linewidth}
			>{\centering\arraybackslash}m{0.09\linewidth}
			>{\centering\arraybackslash}m{0.09\linewidth}
			>{\centering\arraybackslash}m{0.14\linewidth}
			>{\centering\arraybackslash}m{0.14\linewidth}@{}}
			\toprule
			\textbf{Row} & \shortstack{\textbf{Score}\\\textbf{(\%)}} &
			\shortstack{\textbf{Run}\\\textbf{\$}} &
			\shortstack{\textbf{Task}\\\textbf{\$}} & \shortstack{\textbf{Solved}\\\textbf{task \$}} &
			\shortstack{\textbf{Failed}\\\textbf{task \$}} \\
			\midrule
			Plain \codex{}\newline GPT-5 & 67.3/77.9 & 27.71 &
			0.266 & 0.150 & 0.505 \\
			\addlinespace[2pt]
			\mapta{}\newline GPT-5 & 76.9 & 21.38 &
			0.206 & -- & -- \\
			\addlinespace[2pt]
			Plain \codex{}\newline GPT-5.2 & 79.8/88.5 & 50.49 &
			0.486 & 0.247 & 1.429 \\
			\addlinespace[2pt]
			PentestGPT V2\newline GPT-5.2 & 85.0 & -- &
			0.180 & -- & -- \\
			\bottomrule
		\end{tabular}
	\end{table}
	
	The \mapta{} residual is about ten percentage points over the closest
	GPT-5 plain-agent average, so the specialised harness is not merely
	decorative. It also appears more cost-efficient on the fields it
	reports: \$0.206 per challenge, compared with \$0.266 for the plain
	GPT-5 \codex{} baseline. This is a genuine credit to harness design:
	better planning can raise solve rate while reducing wasted exploration.
	\mapta{} does not publish a solve/fail cost split, but the plain-agent
	split shows why that distinction matters: failed \codex{} GPT-5 scans
	cost \$0.505 on average, while successful scans cost \$0.150.
	\pv{} leaves a smaller matched-model residual under GPT-5.2, and its
	reported median per-task cost is below the repriced plain average
	per-task cost. However, that number is not outcome-conditioned: the
	paper does not expose enough aggregate cost fields for the same
	success/failure split.
	
	The other side of the result is coverage. The plain \codex{} P@2
	union slightly exceeds the \mapta{} published score under GPT-5 and
	exceeds the \pv{} GPT-5.2 score by 3.5 percentage points. This does
	not make P@2 identical to a single published run, but it shows that
	repeated plain-agent coverage can match or beat architecture scores on
	challenges solved while still trailing on single-attempt residuals or
	resource efficiency.
	
	\subsection{RQ4: What Does Model Scaling Alone Add?}
	\label{subsec:eval-phase3}
	
	RQ4 keeps the selected \codex{} default-prompt scaffold fixed
	and changes only the model. Table~\ref{tab:model-scaling} reports all
	completed default-prompt model-scaling passes. The pass@1 trend is
	monotonic by model generation: GPT-5 averages 67.3\%, GPT-5.2
	averages 79.8\%, and GPT-5.5 averages 92.3\%.
	
	\begin{table}[H]
		\centering
		\caption{RQ4 model scaling.}
		\label{tab:model-scaling}
		\small
		\setlength{\tabcolsep}{2pt}
		\renewcommand{\arraystretch}{1.16}
		\begin{tabular}{@{}lcccccc@{}}
			\toprule
			\textbf{Model} & \textbf{Cap} & \textbf{P1} &
			\textbf{P2} & \textbf{Avg} & \textbf{P@2} & \textbf{Cost/run} \\
			\midrule
			GPT-5 & \$0.75 & 70 & 70 & 67.3\% & 81 & \$27.71 \\
			GPT-5.2 & \$1.50 & 84 & 82 & 79.8\% & 92 & \$50.49 \\
			GPT-5.5 & \$5.00 & 96 & 96 & 92.3\% & 99 & \$69.53 \\
			\bottomrule
		\end{tabular}
	\end{table}
	
	Table~\ref{tab:tag-results} breaks the same \codex{} model-scaling
	rows down by broad benchmark tag. Tags are multi-label categories:
	a challenge with both IDOR and default-credential behavior contributes
	to both rows, so the table is descriptive rather than a set of
	independent hypothesis tests.
	
	\begin{table}[H]
		\centering
		\caption{Success by benchmark tag.}
		\label{tab:tag-results}
		\small
		\setlength{\tabcolsep}{2pt}
		\renewcommand{\arraystretch}{1.12}
		\begin{tabular}{@{}L{0.40\linewidth} C{0.07\linewidth} C{0.13\linewidth} C{0.13\linewidth} C{0.13\linewidth}@{}}
			\toprule
			\textbf{Tag} & \textbf{$n$} & \textbf{GPT-5} &
			\textbf{GPT-5.2} & \textbf{GPT-5.5} \\
			\midrule
			XSS & 23 & 71.7 & 93.5 & 95.7 \\
			Default Credentials & 18 & 58.3 & 72.2 & 88.9 \\
			IDOR & 15 & 70.0 & 86.7 & 100.0 \\
			Privilege Escalation & 14 & 75.0 & 82.1 & 89.3 \\
			SSTI & 13 & 50.0 & 80.8 & 92.3 \\
			Command Injection & 11 & 68.2 & 86.4 & 100.0 \\
			Business Logic & 7 & 78.6 & 71.4 & 85.7 \\
			Arbitrary File Upload & 6 & 50.0 & 66.7 & 91.7 \\
			Information Disclosure & 6 & 75.0 & 83.3 & 91.7 \\
			Insecure Deserialization & 6 & 50.0 & 66.7 & 83.3 \\
			LFI & 6 & 58.3 & 50.0 & 75.0 \\
			SQLi & 6 & 75.0 & 91.7 & 91.7 \\
			\bottomrule
		\end{tabular}
	\end{table}
	
	The largest GPT-5 to GPT-5.5 gains occur in SSTI, arbitrary file
	upload, insecure deserialization, and command injection. Business
	logic and LFI are less monotonic at GPT-5.2, which is unsurprising
	given their small tag counts and overlapping challenge labels.
	
	GPT-5.5 is not better only because it has a larger cap. If solved
	cells are censored by final per-challenge cost, GPT-5.5 still solves
	89/104 challenges within \$0.75 and 93/104 within \$1.50.
	Table~\ref{tab:resource-usage} also shows that across two passes it
	uses fewer total tokens than GPT-5.2 (102.9M versus 289.1M) while
	solving more unique tasks.
	For context, \pv{} also reports a 91\% headline on the \xbow{} benchmark with
	Opus~4.5 thinking. The plain GPT-5.5 \codex{} row exceeds that
	headline as a pass@1 mean (92.3\%) and under P@2 (95.2\%). This is a
	model-scaling comparison rather than a harness-attribution comparison,
	but it illustrates why architecture claims should be checked against
	current plain-agent baselines.
	
	\subsection{Resource and Cost Analysis}
	
	Tables~\ref{tab:resource-cost} and~\ref{tab:resource-usage} follow
	the style of resource reporting used by \pv{}: each row shows not
	just solve rate, but also cost, tokens, time, and tool use. Every row
	is computed over two completed passes. Table~\ref{tab:resource-cost}
	separates solved scans from failed scans because most wasted budget is
	spent after the agent has not found a viable exploit path. Some agents
	are cheaper but solve fewer tasks; some prompts spend more and solve
	less; GPT-5.5 is expensive per token but unusually token-efficient.
	
	\begin{table}[H]
		\centering
		\caption{Outcome-conditioned scan cost.}
		\label{tab:resource-cost}
		\small
		\setlength{\tabcolsep}{0.8pt}
		\renewcommand{\arraystretch}{1.15}
		\begin{tabular}{@{}L{0.27\linewidth} C{0.11\linewidth} C{0.12\linewidth} C{0.12\linewidth} C{0.13\linewidth} C{0.13\linewidth}@{}}
			\toprule
			\textbf{Run} & \textbf{P@2} & \textbf{Run \$} &
			\textbf{Task \$} & \shortstack{\textbf{Solved}\\\textbf{task \$}} &
			\shortstack{\textbf{Failed}\\\textbf{task \$}} \\
			\midrule
			\codex{} GPT-5 & 77.9\% & 27.71 & 0.266 & 0.150 & 0.505 \\
			\opencode{} GPT-5 & 64.4\% & 11.93 & 0.115 & 0.067 & 0.170 \\
			\piagent{} GPT-5 & 62.5\% & 12.00 & 0.115 & 0.060 & 0.171 \\
			Detailed prompt & 65.4\% & 32.81 & 0.316 & 0.150 & 0.551 \\
			System prompt & 72.1\% & 32.28 & 0.310 & 0.133 & 0.593 \\
			\codex{} GPT-5.2 & 88.5\% & 50.49 & 0.486 & 0.247 & 1.429 \\
			\codex{} GPT-5.5 & 95.2\% & 69.53 & 0.669 & 0.435 & 3.474 \\
			\bottomrule
		\end{tabular}
	\end{table}
	
	\begin{table}[H]
		\centering
		\caption{Token and runtime profile.}
		\label{tab:resource-usage}
		\scriptsize
		\setlength{\tabcolsep}{2pt}
		\renewcommand{\arraystretch}{1.15}
		\begin{tabular}{@{}L{0.34\linewidth} C{0.15\linewidth} C{0.18\linewidth} C{0.13\linewidth} C{0.12\linewidth}@{}}
			\toprule
			\textbf{Run} & \textbf{Tok.} & \textbf{Tok./Solve} &
			\textbf{Tools} & \textbf{Med. s} \\
			\midrule
			\codex{} GPT-5 & 183.7M & 2.27M & 5077 & 165 \\
			\opencode{} GPT-5 & 95.6M & 1.43M & 5259 & 118 \\
			\piagent{} GPT-5 & 67.8M & 1.04M & 3926 & 83 \\
			Detailed prompt & 206.3M & 3.03M & 5351 & 197 \\
			System prompt & 203.5M & 2.71M & 6626 & 189 \\
			\codex{} GPT-5.2 & 289.1M & 3.14M & 6965 & 106 \\
			\codex{} GPT-5.5 & 102.9M & 1.04M & 4718 & 54 \\
			\bottomrule
		\end{tabular}
	\end{table}
	
	The resource tables change the interpretation of several rows.
	\opencode{} and \piagent{} are cheaper than \codex{}, but their
	lower solve counts make them weaker baselines for architecture
	attribution. The security prompts are worse on both axes: they cost
	more and solve fewer tasks than native \codex{}. The solved/failed
	split also shows why failures dominate cost: a successful GPT-5
	\codex{} scan averages \$0.150, but a failed one averages \$0.505;
	for GPT-5.5, the same split is \$0.435 versus \$3.474. GPT-5.5 has
	the highest average scan cost because its price schedule is higher,
	but it is the most token-efficient repeated-run row and has the
	shortest median runtime in the resource tables.
	
	\subsection{Published Resource Anchors}
	
	The comparison papers do not report identical resource metrics, so
	Table~\ref{tab:efficiency} preserves the published values in their
	original form. These rows are not strict efficiency head-to-heads,
	but they give readers the scale of reported operating cost.
	
	\begin{table}[H]
		\centering
		\caption{Published resource anchors.}
		\label{tab:efficiency}
		\small
		\setlength{\tabcolsep}{2pt}
		\renewcommand{\arraystretch}{1.15}
		\begin{tabular}{@{}L{0.27\linewidth} L{0.20\linewidth} L{0.39\linewidth}@{}}
			\toprule
			\textbf{System} & \textbf{\xbow{} Benchmark Result} &
			\textbf{Reported Resource Use} \\
			\midrule
			\mapta{} & 80/104 (76.9\%) &
			\$21.38 total; \$0.206 average per challenge; 143.2s median solve time \\
			\pv{} & 85\% with GPT-5.2 thinking &
			12 median LLM calls, 3.2 minutes, and \$0.18 median cost per \xbow{} benchmark task \\
			\bottomrule
		\end{tabular}
	\end{table}
	
	\section{Analysis}
	\label{sec:analysis}
	
	\subsection{Architecture Residuals}
	
	For each comparison system $s$, let $P_s$ be the published solve
	rate of the architecture-heavy system and $C_s$ be the closest
	model-matched plain-agent solve rate. We define the architecture
	residual as:
	\begin{equation}
		\Delta_s = P_s - C_s .
	\end{equation}
	A positive residual means the purpose-built system solves more than
	the plain agent under the closest available model match. A small
	residual suggests that much of the result may already be explained
	by the model plus a generic coding-agent scaffold.
	
	\begin{table}[H]
		\centering
		\caption{Residual interpretation.}
		\label{tab:residual-interpretation}
		\small
		\setlength{\tabcolsep}{4pt}
		\renewcommand{\arraystretch}{1.15}
		\begin{tabular}{@{}l L{0.58\linewidth}@{}}
			\toprule
			\textbf{Observed Gap} & \textbf{Interpretation} \\
			\midrule
			$\Delta_s \leq 0$ &
			Plain coding agent matches or exceeds the specialised system. \\
			$0 < \Delta_s \leq 5$ pp &
			Systems are practically comparable; architecture adds little
			benchmark-visible value. \\
			$5 < \Delta_s \leq 15$ pp &
			Architecture may help; category and trace analysis are needed. \\
			$\Delta_s > 15$ pp &
			Specialised architecture provides a substantial matched-model
			advantage. \\
			\bottomrule
		\end{tabular}
	\end{table}
	
	Under this scale, \mapta{} shows a meaningful residual over GPT-5
	\codex{}, while \pv{} shows a smaller residual over GPT-5.2
	\codex{}. In both cases, however, the plain-agent P@2 union reaches or
	exceeds the published score. The residual and union therefore answer
	different questions: architecture appears to improve single-attempt
	efficiency, while repeated plain-agent coverage can close or exceed
	the challenge-count gap. The GPT-5.5 row is reserved for model-scaling
	analysis rather than architecture attribution.
	
	\subsection{Variance and Repeatability}
	
	The two-pass results show why a single pass should not be
	overinterpreted. The default \codex{} GPT-5 runs both solve 70
	challenges, but only 59 challenges are solved in both passes.
	Another 22 challenges flip between solved and unsolved across the
	two attempts. This is why the paper reports pass@1 for
	single-attempt capability, reliable solves for repeatability, and
	pass@2 for stochastic coverage.
	
	\subsection{Where Architecture May Be Adding Value}
	
	The residual gaps are most likely to come from mechanisms that a
	plain coding CLI does not have by design:
	\begin{itemize}
		\item \textbf{State tracking:} long chains involving credentials,
		cookies, and privilege transitions may benefit from structured
		memory.
		\item \textbf{Search control:} attack-tree search or difficulty
		estimation can stop an agent from overcommitting to a weak path.
		\item \textbf{Verification:} a separate validation loop can demand
		exploit evidence rather than plausible narration.
		\item \textbf{Browser-heavy tasks:} DOM state, JavaScript execution,
		and visual confirmation can benefit from first-class browser
		automation.
	\end{itemize}
	This interpretation is consistent with \pv{}'s emphasis on
	difficulty-aware planning and resource analysis, but our result adds
	a baseline requirement: those mechanisms should be compared against
	a strong plain coding agent, not only against weaker chat-style or
	tool-poor baselines.
	
	\subsection{Reproducibility Checklist}
	
	Future architecture papers should report:
	\begin{itemize}
		\item exact model name, inference settings, and access date;
		\item benchmark commit, excluded tasks, and local repairs;
		\item pass@1, repeated-pass variance, and pass@2 where available;
		\item wall-clock, token, tool-call, budget/cap, and
		success/failure cost statistics;
		\item a model-matched plain coding-agent baseline;
		\item enough per-challenge traces to support failure analysis.
	\end{itemize}
	Without these controls, a headline score can conflate model
	progress, extra budget, prompt changes, and architecture design.
	
	\section{Discussion}
	\label{sec:discussion}
	
	\subsection{Simple Takeaway}
	
	The answer is not that harnesses are useless and not that bare agents
	are enough. The data support a middle position. A plain \codex{} CLI
	is already strong: across two-pass default-prompt runs, it averages
	67.3\% pass@1 on GPT-5, 79.8\% on GPT-5.2, and 92.3\% on GPT-5.5,
	with unions rising from 81/104 (77.9\%) to 92/104 (88.5\%) and then
	99/104 (95.2\%). However, \mapta{} and \pv{} still show positive
	matched-model residuals, and \mapta{} reports lower cost per challenge
	than the plain GPT-5 baseline. Architecture can therefore add
	measurable value even when repeated plain-agent coverage is high.
	
	\subsection{Prompting Is Not a Harness}
	
	The custom prompt rows are a useful negative control. If security
	prompting alone were enough, the detailed prompt or whole-system
	prompt should have improved \codex{}. Instead, both completed rows
	underperform the native default and cost more. Prompt-heavy designs
	should therefore be compared against the unmodified agent default
	before being credited as architecture.
	
	\subsection{Model Scaling Can Hide Harness Effects}
	
	The GPT-5.5 row shows why model-matched baselines matter. A plain
	agent with a newer model can outperform older purpose-built headlines
	even without new security machinery: here, GPT-5.5 \codex{} exceeds
	\pv{}'s 91\% Opus~4.5 headline. That does not invalidate harness
	design, but it means harness claims need to be separated from model
	progress.
	
	\subsection{Threats to Validity}
	
	\noindent\textbf{Cross-study attribution.}
	\mapta{} and \pv{} are not rerun inside our infrastructure. Their rows
	are therefore descriptive comparisons against published results, not
	direct architectural ablations.
	
	\noindent\textbf{Model--budget confounding and order.}
	RQ4 changes the model and the cost cap together, and the model runs
	were conducted sequentially rather than randomized. Budget-censored
	analysis helps, but it cannot fully reproduce the trajectory of a
	prospectively lower-cap run.
	
	\noindent\textbf{GPT-5.2 repricing.}
	GPT-5.2 runs were live-stopped using the experiment logger's
	then-configured placeholder estimate. The paper reports post-hoc
	official-price costs from the recorded token counts, so GPT-5.2 dollar
	columns are appropriate for resource accounting but not identical to
	rerunning the same traces under a prospectively enforced official-price
	cap.
	
	\noindent\textbf{Two trials.}
	Two passes reveal substantial run-to-run flips, but they are still
	insufficient for precise per-task probability estimates. Additional
	repeated trials would tighten confidence intervals, especially for the
	prompt rows.
	
	\noindent\textbf{Public-benchmark contamination.}
	Random flags prevent literal flag reuse, but they do not rule out prior
	knowledge of public challenge structure. Held-out targets or
	semantics-preserving challenge mutations would measure generalization
	more directly.
	
	\noindent\textbf{Benchmark repairs.}
	Approximately 40 targets required local image or build repairs. The
	repairs were necessary to run the full validation set, but they make
	our fork not byte-identical to cross-paper benchmark checkouts.
	
	\noindent\textbf{CLI--model coupling.}
	All plain-agent CLI comparisons use GPT-series models. This may favor
	\codex{} because it is designed around OpenAI models and may expose a
	more mature integration path for them than the other CLIs. A broader
	replication should include non-GPT models across all three agents.
	
	\noindent\textbf{Scope and egress.}
	The \xbow{} benchmark is a standardized CTF-style web benchmark, not a full
	real-world penetration test with reconnaissance, reporting, active
	defenses, or multiple hosts. The jail blocks host and Docker access,
	but released aggregate data alone cannot independently audit every
	network destination the runtime attempted.
	
	\section{Ethical Considerations}
	\label{sec:ethics}
	
	\noindent\textbf{Dual-use responsibility.}
	Autonomous penetration testing research is dual-use. This paper does
	not introduce a new exploit framework or a new attack architecture; it
	evaluates existing general-purpose coding agents on an intentionally
	vulnerable benchmark. We publish the methodology because clear
	baselines help the community assess capability claims more responsibly.
	Any use of these methods outside benchmark environments should occur
	only in authorised security testing with explicit written scope.
	
	\medskip
	\noindent\textbf{Controlled evaluation.}
	All experiments were conducted exclusively against purpose-built,
	isolated Docker containers provided by the \xbow{} benchmark. No
	production systems, third-party infrastructure, or live services were
	tested at any point during this research.
	
	\medskip
	\noindent\textbf{Responsible disclosure.}
	No previously undisclosed vulnerabilities were discovered. The released
	artifact includes benchmark harness code, configuration files,
	aggregate results, per-attempt tables, and a small set of redacted
	representative traces. Full raw logs are withheld to avoid
	distributing unnecessary exploit payload collections beyond what is
	required for reproducibility.
	
	\section{Conclusion}
	\label{sec:conclusion}
	
	This paper argues for baseline-first evaluation of autonomous
	penetration-testing agents. On the \xbow{} benchmark, the default \codex{} CLI is a
	strong plain-agent baseline: it solves 70/104 tasks in each of its
	first two GPT-5 passes and 81/104
	(77.9\%) under a two-pass union. \opencode{} and \piagent{} trail
	behind in the same matched setting. The two tested security-prompt
	variants do not improve the observed \codex{} score in this experiment.
	
	The architecture comparisons show a more nuanced picture than either
	``the harness does everything'' or ``the harness does not matter.''
	\mapta{} reports 76.9\% with GPT-5, compared with a 67.3\%
	two-pass \codex{} GPT-5 pass@1 average, leaving a 9.6-point residual
	under unequal effort and budget settings. It also reports lower
	average cost per challenge than the plain GPT-5 baseline. With
	GPT-5.2, the residual against \pv{}'s GPT-5.2-thinking result shrinks
	to about 5.2 points, and the \codex{} two-pass union exceeds that
	published target. With GPT-5.5, the same plain scaffold solves 96/104
	tasks in both passes and 99/104 (95.2\%) under P@2, exceeding \pv{}'s
	91\% Opus~4.5 headline as a model-scaling result.
	
	The broader lesson is therefore methodological. Security harnesses can
	add benchmark lift and may improve cost efficiency, but their
	contribution should be measured against strong, model-matched
	plain-agent baselines and separated from backbone model progress.
	

	
	\appendices
	
	\section{Reproducibility Artifact}
	\label{app:artifact}
	
	The accompanying GitHub artifact,
	\url{https://github.com/krodalabs/coding-agent-research-artifact},
	contains the material needed to inspect or regenerate the reported
	results: exact prompts, recorded agent configurations, the benchmark
	repair patch, harness code, the canonical 1,456-row per-attempt table,
	its data dictionary, analysis scripts, generated paper metrics, and two
	flag-redacted representative traces.
	
\end{document}